\documentclass[draft,showkeys,showpacs,eqsecnum,nofootinbib]{revtex4}
\renewcommand{\theequation}{\arabic{equation}}
\def\beq{\begin{equation}}
\def\eeq{\end{equation}}
\def\bea{\begin{eqnarray}}
\def\eea{\end{eqnarray}}\def\nn{\nonumber}
\def\pr{\prime}
\def\na{\nabla}

\def\nn{\nonumber}

\begin{document}
\title{Warp products and (2+1) dimensional spacetimes}
\author{Soon-Tae Hong}
\affiliation{Department of Science
Education and Research Institute for Basic Sciences, Ewha Womans
University, Seoul 120-750, Republic of Korea}
\author{Yeji Kim}
\affiliation{Department of Physics, 
Ewha Womans University, Seoul 120-750, Republic of Korea}
\date{\today}
\begin{abstract}
We investigate the geometrical aspects of the extended (2+1) dimensional Banados-Teitelboim-Zanelli 
spacetimes in the multiply warped product scheme. To do this, we analyze the interior physical 
properties by constructing the explicit warp functions in these regions. 
\end{abstract}
\maketitle

%%%%%%%%%%%%%%%%%%%%%%%%%%%%%%%%%%%%%%%%%%%%%%%%%%%%%%%%%%%%%%%%%%%%%%%%
\section{Introduction}
\setcounter{equation}{0}
\renewcommand{\theequation}{\arabic{section}.\arabic{equation}}
%%%%%%%%%%%%%%%%%%%%%%%%%%%%%%%%%%%%%%%%%%%%%%%%%%%%%%%%%%%%%%%%%%%%%%%%

Since the (2+1) dimensional Banados-Teitelboim-Zanelli (BTZ)~\cite{btz1,btz2,btz3} 
black hole has been proposed, significant interests in this black hole have 
been aroused with the novel discovery that the thermodynamics of higher dimensional 
black holes can be interpreted in terms of the BTZ solutions~\cite{mal98,sfe98}. There have 
been also tremendous advances in lower dimensional black holes related with the string 
theory since an exact conformal field theory describing a black hole in the (1+1) 
dimensional spacetime was proposed~\cite{witten91}. The BTZ black hole has been also generalized to 
possess the scalar tensor (ST) theories~\cite{lemos95,sa96,chan97,pimentel89,hong002}. The Hawking and Unruh effects of the 
(2+1) dimensional black holes have been analyzed in terms of the global embedding Minkowski 
space (GEMS) approach~\cite{deser97,deser98,deser99} to produce the novel global higher dimensional 
flat embeddings of the (2+1) dimensional static, rotating and charged BTZ black holes and de Sitter spacetimes~\cite{hong00}. The (2+1) dimensional ST theories have been later analyzed~\cite{hong002} in terms of the GEMS scheme.

Now, we consider the ST theories in terms of the Lagrangian associated with scalar fields. To do this, we consider an action for the ST theories~\cite{pimentel89,chan97,hong002}:
\beq
S=\int~d^{3}x~\sqrt{-g}~[c(\chi)R-\omega(\chi)(\na \chi)^{2}+V(\chi)],
\label{action0}
\eeq
where $R$ is a scalar curvature, $c(\chi)$ and $\omega(\chi)$ are coupling functions for 
the scalar field $\chi$ and $V(\chi)$ is a potential function, respectively. For instance, the choce of 
$c(\chi)=1$, $\omega(\chi)=0$ and $V(\chi)=2\Lambda$ corresponds to the static BTZ black hole solution.
For the case of $c(\chi)=\chi$, we have the ST theories~\cite{pimentel89,chan97,hong002}. In (2+1)-dimensions, we take an ansatz for 
three-metric of the ST theories:
\beq
ds^{2}=-f(r)dt^{2}+\frac{1}{f(r)}dr^{2}+r^{2}d\theta^{2}.
\label{stmetric3}
\eeq
Using the Einstein equation obtained from the action in (\ref{action0}) and the metric in (\ref{stmetric3}), we obtain~\cite{chan97}
\bea
f^{\pr\pr}+\frac{f^{\pr}}{r}&=&\frac{2V}{\chi}-\frac{1}{\chi}\left[f^{\pr}\chi^{\pr}+\frac{2}{r}(fr\chi^{\pr})^{\pr}\right],
\label{fff1}\\
\chi^{\pr\pr}+\omega(\chi^{\pr})^{2}&=&0,\label{fff2}\\
f^{\pr}r&=&\frac{r^{2}V}{\chi}-\frac{1}{\chi}(fr^{2}\chi^{\pr})^{\pr},\label{fff3}
\eea
where the prime denotes ordinary derivative with respect to $r$. We next assume special forms of $V(\chi)$ and $\chi$ as in 
(\ref{lagb}) and (\ref{chidef}), respectively. From (\ref{fff2}), we fix $\omega$, and from (\ref{fff1}) and (\ref{fff3}) we 
obtain $f(r)$. Here one notes that in this approach one cannot specify uniquely $V(\chi)$ or $\chi$~\cite{chan97}. 

In this paper, we will consider two different cases with $c(\chi)=\chi$ where gravitational forces are given by a mixture of the metric and the scalar field. Here we note that the scalar field will be just an ordinary one, not a phantom field. 
Exploiting the Lagrangian for the metric and scalar field, we will investigate the Einstein equations and warp product aspects of the ST theories in detail below. To do this, we will exploit an algorithm for roots of cubic and quartic equations associated with the warp products.
  
On the other hand, the warped product spacetimes~\cite{bishop69,beem96} have been attractive since they 
include classical examples of spacetime such as the Friedmann-Robertson-Walker manifold and 
the intermediate zone of the Reissner-Nordstr\"om (RN) black hole
~\cite{rn1,rn2,demers95,ghosh95,cognola98}. The interior 
Schwarzschild spacetime has been represented as a multiply warped spacetime with warping 
functions~\cite{choi00} to produce the Ricci curvature in terms of $f_{1}$ and $f_{2}$ for the 
multiply warped products of the form $M=R\times_{f_1}R\times_{f_2} S^{2}$. The interior 
RN-AdS spacetime has been also investigated by using the multiply warped product scheme~\cite{hong05}.
In the (2+1) dimensional spacetime, the multiply warped product has been studied to investigate 
the interior solutions of the (2+1) BTZ black holes and the exterior solutions of the (2+1) de Sitter 
black holes~\cite{hong03}. In order to study nonsmooth curvatures associated with multiple discontinuities involved in the evolution of the universe, the multiply warped product has been also applied to the Friedmann-Robertson-Walker model~\cite{hong04}. 

In this paper, in order to investigate the physical properties inside the outer event horizon, we 
will analyze the multiply warped product manifold associated
with the ST theories of the BTZ black holes containing the terms proportional to $1/r$ and $1/r^{2}$ as in (\ref{dsbar1}) and (\ref{dsbar2}), respectively. From now on we name them type-I and 
type-II, respectively.  Exploiting the multiply warped product scheme,
we will investigate the interior solutions for multiply warped functions in the (2+1) extended BTZ black holes, 
to explicitly obtain the Ricci and Einstein curvatures inside the outer event horizons of these metrics.

%%%%%%%%%%%%%%%%%%%%%%%%%%%%%%%%%%%%%%%%%%%%%%%%%%%%%%%%%%%%%%%%%%%%%%
\section{Type-I extended BTZ spacetime}
\setcounter{equation}{0}
\renewcommand{\theequation}{\arabic{section}.\arabic{equation}}
%%%%%%%%%%%%%%%%%%%%%%%%%%%%%%%%%%%%%%%%%%%%%%%%%%%%%%%%%%%%%%%%%%%%%%

Now, we consider the Lagrangian associated with an asymptotically constant 
scalar field~\cite{chan97}
\beq
{\cal L}=\sqrt{-g}\left[\chi R-\frac{2}{1-\chi}(\na \chi)^{2}+2(3-3\chi+\chi^{2})\Lambda \chi
+\frac{8M}{27B^{2}}(1-\chi)^{3}\right],
\label{lagb}
\eeq
where
\beq
\chi=\frac{r}{r-3B/2},\label{chidef}
\eeq
whose solution is given by 
\bea
ds^{2}&=&-\bar{N}^{2}dt^{2}+\bar{N}^{-2}dr^{2}+r^{2}d\phi^{2},\nn\\
\bar{N}^{2}&=&-M+\frac{MB}{r}+\frac{r^{2}}{l^{2}},\label{dsbar1}
\eea 
with $l^{2}=1/\Lambda$. Here $M$ is the positive mass parameter, $B$ in (\ref{dsbar1}) is introduced to couple 
the scalar field $\chi$ to the metric $g_{\mu\nu}$ gravity and in that sense $B$ is meaningful. 
On the other hand, investigating the lapse functions $\bar{N}^{2}$ and $N^{2}$ (for warp product case), one 
readily observe that the only curvature singularity is located at $r=0$~\cite{chan97}. 
One notes that the metric (\ref{dsbar1}) looks like the Schwarzschild-AdS metric. If $\Lambda$ vanishes, the 
metric becomes exactly the same form as the four-dimensional Schwarzschild case.
Variating the Lagrangian in (\ref{lagb}) with respect to the metric, we obtain 
Einstein equation
\beq
R_{\mu\nu}-\frac{1}{2}g_{\mu\nu}R=T_{\mu\nu},\label{einstein}
\eeq
where the energy-stress tensor is given by
\beq
T_{\mu\nu}=\frac{2}{\chi(1-\chi)}\na_{\mu}\chi\na_{\nu}\chi+\frac{1}{2}g_{\mu\nu}
\left[-\frac{2}{\chi(1-\chi)}(\na \chi)^{2}+2(3-3\chi+\chi^{2})\Lambda+\frac{8M}{27B^{2}\chi}(1-\chi)^{3}\right].
\label{tmunu}
\eeq
Here we reemphasize that the scalar field is an ordinary one, not a phantom field. Moreover, one can readily check that, 
for the case of warp product with $N^{2}$ in (\ref{btzlapse}), we have the same Einstein equation with the same 
energy-stress tensor (\ref{tmunu}). 
Next, variation of the Lagrangian in (\ref{lagb}) with respect to the scalar field $\chi$ yields
\beq
R-\frac{2}{(1-\chi)^{2}}(\na\chi)^{2}+\frac{4}{1-\chi}\na^{2}\chi+6(1-\chi)^{2}-\frac{8M}{9B^{2}}(1-\chi)^{2}=0.
\eeq 

%%%%%%%%%%%%%%%%%%%%%%%%%%%%%%%%%%%%%%%%%%%%%%%%%%%%%%%%%%%%%%%%%%%%%%
\subsection{Static case}
%%%%%%%%%%%%%%%%%%%%%%%%%%%%%%%%%%%%%%%%%%%%%%%%%%%%%%%%%%%%%%%%%%%%%%

Now, in order to investigate physical properties inside the black hole event horizon, we
consider a multiply warped product manifold whose metric is of the form~\cite{choi00,unal00,hong05,hong03}
\beq
g=-d\mu^{2}+\sum_{i=1}^{n}f_{i}^{2}g_{i},
\label{gnew}
\eeq
where $f_{i}$ are positive warping functions and $g_{i}$ are the corresponding metrics, 
to extend the warped product spaces to richer class of spaces involving multiply
products. In order to construct a multiply warped product manifold for
the static type-I extended BTZ interior solution, we start
with the three-metric inside the event horizon
\beq
ds^{2}=N^{2}dt^{2}-N^{-2}dr^{2}+r^{2}d\phi^{2}
\label{btzmetric}
\eeq
with the lapse function for the interior solution
\beq
N^{2}=M-\frac{MB}{r}-\frac{r^{2}}{l^{2}}.
\label{btzlapse}
\eeq
Here the parameters $M$, $B$ and $l$ are the same as those in (\ref{dsbar1}). However, these parameters in (\ref{btzlapse}) 
are defined inside the event horizon of the extended BTZ black hole. 
The roots of $N^{2}=0$ for arbitrary real value of $r$ are given by
\bea
r&=&r_{1},\nn\\
r&=&r_{2}=\frac{1}{2}\left[-r_{1}+\left(r_{1}^{2}+\frac{4Ml^{2}B}{r_{1}}\right)^{1/2}\right]=-r_{4},\nn\\
r&=&r_{3}=\frac{1}{2}\left[-r_{1}-\left(r_{1}^{2}+\frac{4Ml^{2}B}{r_{1}}\right)^{1/2}\right]=-r_{5},
\label{3roots}\eea
which shows that we do not have three positive roots. 
Here one notes that, since $r$ is a radial distance, we have a constraint about $r$, namely $r>0$, 
if we exclude the curvature singularity at $r=0$ in the range of $r$. We also observe that for instance 
$r_{3}$ is negative for $r_{1}>0$ and thus we will use the factor $(r+r_{5})~(r_{5}>0)$ instead of $(r-r_{3})$ to emphasize 
that $r_{3}$ is not a positive real root. We obtain an identity
\beq
r_{1}+r_{2}+r_{3}=0.\label{r1r2r3}
\eeq
Moreover, for a given nonzero $B$, the mass parameter $M$ is given in terms 
of the three roots as follows
\beq
M=-\frac{r_{1}r_{2}r_{3}}{l^{2}B}.
\eeq
For a vanishing $B$, we also obtain $M$ in terms of $r_{1}$ in (\ref{mbzero}) below. 
The lapse function can be now categorized in terms of the value of $B$ and $r_{1}$.

(i) For $0<B<B_{1}$ with 
\beq
B_{1}=\frac{2r_{1}^{3}}{Ml^{2}}
\eeq
and $r_{1}>0$ we find
\beq
r_{1}<\left(r_{1}^{2}+\frac{4Ml^{2}B}{r_{1}}\right)^{1/2}<3r_{1},\label{sqrtroot},
\eeq
and thus we obtain two positive real roots $r_{1}$ and $r_{2}$ with $r_{1}>r_{2}$ to yield
\beq
N^{2}=\frac{(r_{1}-r)(r-r_{2})(r+r_{5})}{l^{2}r}
\label{lapserh1}
\eeq
which is well-defined in the region $r_{2}<r<r_{1}$. Now we define a new coordinate $\mu$ as follows
\beq
d\mu^{2}=N^{-2}dr^{2},
\label{btzdmu}
\eeq
which can be integrated to yield
\beq
\mu=\int_{r_{2}}^{r}dx~\left(\frac{l^{2}x}{(r_{1}-x)(x-r_{2})(x+r_{5})}\right)^{1/2},
\label{btzdmu11}
\eeq
whose analytic solution is of the form
\beq
\mu=\frac{2r_{2}l}{[(r_{2}+r_{5})r_{1}]^{1/2}}\Pi\left(c_{11},c_{12},c_{13}\right)=G(r).
\label{btzsolmu}
\eeq
Here the arguments $(c_{11},c_{12},c_{13})$ are given by
\bea
c_{11}&=&\sin^{-1}\left(\frac{(r-r_{2})r_{1}}{(r_{1}-r_{2})r}\right)^{1/2},\nn\\
c_{12}&=&\frac{r_{1}-r_{2}}{r_{1}},\nn\\
c_{13}&=&\left(\frac{(r_{1}-r_{2})r_{5}}{(r_{5}+r_{2})r_{1}}\right)^{1/2},
\eea
and $\Pi(a,b,c)$ is the elliptic integral of the third kind defined as~\cite{table}
\beq
\Pi(a,b,c)=\int_{0}^{a}\frac{dx}{(1-b\sin^{2}x)(1-c^{2}\sin^{2}x)^{1/2}}.
\label{piabc}
\eeq
Moreover, $G(r)$ in (\ref{btzsolmu}) satisfies the following boundary condition
\beq
{\rm lim}_{r\rightarrow r_{2}}G(r)=0.
\label{btzbdy1}
\eeq
One notes that $dr/d\mu >0$ implies that $G^{-1}$ is a well-defined function.

(ii) For $B\geq B_{1}$ and $r_{1}>0$ we find two positive real roots $r_{1}$ and $r_{2}$ with $r_{1}\leq r_{2}$ to yield
\beq
N^{2}=\frac{(r_{2}-r)(r-r_{1})(r+r_{5})}{l^{2}r}
\label{lapserh2}
\eeq
which is well-defined in the region $r_{1}<r<r_{2}$. The remnant arguments are the same as those of the case (i), 
if we exchange $r_{1}$ for $r_{2}$ as in (\ref{lapserh1}) and (\ref{lapserh2}). We thus arrive at 
\beq
\mu=\frac{2r_{1}l}{[(r_{1}+r_{5})r_{2}]^{1/2}}\Pi\left(c_{21},c_{22},c_{23}\right)=G(r).
\label{btzsolmu2}
\eeq 
where
\bea
c_{21}&=&\sin^{-1}\left(\frac{(r-r_{1})r_{2}}{(r_{2}-r_{1})r}\right)^{1/2},\nn\\
c_{22}&=&\frac{r_{2}-r_{1}}{r_{2}},\nn\\
c_{23}&=&\left(\frac{(r_{2}-r_{1})r_{5}}{(r_{5}+r_{1})r_{2}}\right)^{1/2}.
\eea
Now, $G(r)$ in (\ref{btzsolmu}) fulfills the boundary condition
\beq
{\rm lim}_{r\rightarrow r_{1}}G(r)=0,
\label{btzbdy122}
\eeq
and $dr/d\mu >0$ indicates that $G^{-1}$ is a well-defined function.

(iii) For $B=0$ and $r_{1}>0$, which corresponds to the BTZ black hole, 
we find a positive real root $r_{1}$ to yield
\beq
N^{2}=\frac{(r_{1}-r)(r+r_{1})}{l^{2}}
\label{lapserh3}
\eeq
where 
\beq
r_{1}=M^{1/2}l,
\eeq
which is well-defined in the region $0<r<r_{1}$. We also obtain $M$ as a function of $r_{1}$:
\beq
M=\frac{r_{1}^{2}}{l^{2}}.\label{mbzero}
\eeq
Now we define the coordinate $\mu$ in (\ref{btzdmu})
which can be integrated to yield
\beq
\mu=\int_{0}^{r}dx~\left(\frac{l^{2}}{(r_{1}-x)(x+r_{1})}\right)^{1/2},
\label{btzdmu130}
\eeq
whose analytic solution is of the form
\beq
\mu=l\sin^{-1}\frac{r}{r_{1}}=G(r).
\label{btzdmu13}
\eeq
This form is consistent with the previous work for the BTZ spacetime~\cite{hong03}. 
Here $G(r)$ in (\ref{btzdmu13}) satisfies the boundary condition
\beq
{\rm lim}_{r\rightarrow 0}G(r)=0
\label{btzbdy3}
\eeq
and $dr/d\mu >0$ implies that $G^{-1}$ is a well-defined function.

(iv) For $-B_{2}<B<0$ with 
\beq
B_{2}=\frac{r_{1}^{3}}{4Ml^{2}}
\eeq
and $r_{1}>0$ we find
\beq
0<\left(r_{1}^{2}+\frac{4Ml^{2}B}{r_{1}}\right)^{1/2}<r_{1},\label{sqrtroot4}
\eeq
and thus we obtain a positive real root $r_{1}$ and two negative real roots $r_{2}$ and $r_{3}$ to produce
\beq
N^{2}=\frac{(r_{1}-r)(r+r_{4})(r+r_{5})}{l^{2}r},
\label{lapserh4}
\eeq
which is well-defined in the region $0<r<r_{1}$. Now we define the coordinate $\mu$ in (\ref{btzdmu})
which can be integrated to yield
\beq
\mu=\int_{0}^{r}dx~\left(\frac{l^{2}x}{(r_{1}-x)(x+r_{4})(x+r_{5})}\right)^{1/2},
\label{btzdmu14}
\eeq
whose analytic solution is given by 
\beq
\mu=\frac{2r_{4}l}{[(r_{1}+r_{4})r_{5}]^{1/2}}\left[\Pi(c_{31},c_{32},c_{33})-F(c_{31},c_{33})\right]=G(r).
\label{btzdmu142}
\eeq
Here the arguments $(c_{31},c_{32},c_{33})$ are found to be
\bea
c_{31}&=&\sin^{-1}\left(\frac{(r_{1}+r_{4})r}{r_{1}(r+r_{4})}\right)^{1/2},\nn\\
c_{32}&=&\frac{r_{1}}{r_{1}+r_{4}},\nn\\
c_{33}&=&\left(\frac{r_{1}(r_{5}-r_{4})}{(r_{1}+r_{4})r_{5}}\right)^{1/2}.
\eea 
Moreover, $\Pi(a,b,c)$ is given by (\ref{piabc}) and $F(a,b)$ is the elliptic integral of the first kind defined as~\cite{table}
\beq
F(a,b)=\int_{0}^{a}\frac{dx}{(1-b^{2}\sin^{2}x)^{1/2}}.
\label{piabc22}
\eeq
Moreover, $G(r)$ in (\ref{btzdmu14}) satisfies the boundary condition
(\ref{btzbdy3}) and $dr/d\mu >0$ implies that $G^{-1}$ is a well-defined function.

(v) For $B<-B_{2}$ and $r_{1}>0$ we find
\beq
\left(r_{1}^{2}+\frac{4Ml^{2}B}{r_{1}}\right)^{1/2}<0,\label{sqrtroot5}
\eeq
and thus we obtain a positive real root $r_{1}$ and two imaginary roots $r_{2}$ and $r_{3}$ to produce
\beq
N^{2}=\frac{(r_{1}-r)\left(r^{2}+r_{1}r-\frac{Ml^{2}B}{r_{1}}\right)}{l^{2}r}.
\label{lapserh5}
\eeq
Now we define the coordinate $\mu$ in (\ref{btzdmu})
which can be integrated to yield
\beq
\mu=\int_{0}^{r}dx~\left(\frac{l^{2}x}{(r_{1}-x)\left(x^{2}+r_{1}x-\frac{Ml^{2}B}{r_{1}}\right)}\right)^{1/2}=G(r),
\label{btzdmu15}
\eeq
which is well-defined in the region $0<r<r_{1}$. Here, $G(r)$ in (\ref{btzdmu15}) fulfills the boundary condition
(\ref{btzbdy3}) and $dr/d\mu >0$ implies that $G^{-1}$ is a well-defined function. If a root of $N^{2}=0$ does not belong to the above 
categories, we cannot construct the coordinate $\mu$.

Exploiting the coordinate $\mu$ in (\ref{btzsolmu}), (\ref{btzsolmu2}), (\ref{btzdmu13}), (\ref{btzdmu142}) and (\ref{btzdmu15}), we rewrite the
metric (\ref{btzmetric}) as warped products
\beq
ds^{2}=-d\mu^{2}+f_{1}(\mu)^{2}dt^{2}+f_{2}^{2}(\mu)d\phi^{2}
\label{btzmetric2}
\eeq
where
\bea
f_{1}(\mu)&=&\left(M-\frac{MB}{G^{-1}(\mu)}-\frac{(G^{-1}(\mu))^{2}}{l^{2}}\right)^{1/2},
\nonumber\\
f_{2}(\mu)&=&G^{-1}(\mu).
\label{btzf1f2}
\eea

After some algebra using (\ref{btzsolmu}) and the warp products in (\ref{btzmetric2}), 
we obtain the following nonvanishing Ricci curvature components
\bea
R_{\mu\mu}&=&-\frac{f_{1}^{\pr\pr}}{f_{1}}-\frac{f_{2}^{\pr\pr}}{f_{2}},
\nonumber\\
R_{tt}&=&\frac{f_{1}f_{1}^{\pr}f_{2}^{\pr}}{f_{2}}+f_{1}f_{1}^{\pr\pr},
\nonumber\\
R_{\phi\phi}&=&\frac{f_{1}^{\pr}f_{2}f_{2}^{\pr}}{f_{1}}+f_{2}f_{2}^{\pr\pr}.
\label{btzricci}
\eea
Using the explicit expressions for $f_{1}$ and $f_{2}$ in (\ref{btzf1f2}), one obtains
the Ricci curvature components
\bea
R_{\mu\mu}&=&-\frac{2f_{1}^{\pr}}{f_{2}}+\frac{3MB}{2f_{2}^{3}},\nonumber\\
R_{tt}&=&\frac{2f_{1}^{2}f_{1}^{\pr}}{f_{2}}-\frac{3MBf_{1}^{2}}{2f_{2}^{3}},\nonumber\\
R_{\phi\phi}&=&2f_{2}f_{1}^{\pr},
\label{btzriccis}
\eea
and the Einstein scalar curvature
\beq
R=-\frac{6}{l^{2}},
\label{btzeinr}
\eeq
in the interior of the outer event horizon of the static type-I extended BTZ black hole.
 
%%%%%%%%%%%%%%%%%%%%%%%%%%%%%%%%%%%%%%%%%%%%%%%%%%%%%%%%%%%%%%%%%%%%%%%
\subsection{Rotating case}
%%%%%%%%%%%%%%%%%%%%%%%%%%%%%%%%%%%%%%%%%%%%%%%%%%%%%%%%%%%%%%%%%%%%%%

Now we consider a multiply warped product manifold associated with the
rotating type-I extended BTZ black hole inside the event horizon whose three-metric is given by
\beq
ds^{2}=N^{2}dt^{2}-N^{-2}dr^{2}+r^{2}(d\phi+N^{\phi}dt)^{2}
\label{rotbtzmetric}
\eeq
where the lapse and shift functions are found to become
\bea
N^{2}&=&M-\frac{MB}{r}-\frac{r^{2}}{l^{2}}-\frac{J^{2}}{4r^{2}},
\label{rotbtzlapse1}
\nonumber\\
N^{\phi}&=&-\frac{J}{2r^{2}},
\label{rotbtzlapse2}
\eea
with an angular momentum $J$.  Note that four roots of the equation
$N^{2}=0$ yield the lapse function in terms of the event horizons
as follows
\beq
N^{2}=\frac{(r_{+}-r)(r-r_{-})(r-r_{1})(r-r_{2})}{l^{2}r^{2}}
\label{rotlapserh}
\eeq
which, for the interior solution, is well-defined in the region $r_{-}<r<r_{+}$.
The four roots $(r_{+},r_{-},r_{1},r_{2})$ of $N^{2}=0$ together with $r_{2}<r_{1}<r_{-}<r_{+}$ 
are analyzed in Appendix A. Because classification of roots of a quartic equation is highly nontrivial, 
we impose a restriction that $r_{-}>0$ to proceed to obtain the coordinate $\mu$.

Defining the coordinate $\mu$ as in (\ref{btzdmu}), we obtain
\beq
\mu=\int_{r_{-}}^{r}dx~\left(\frac{l^{2}x^{2}}{(r_{+}-x)(x-r_{-})(x-r_{1})(x-r_{2})}\right)^{1/2},
\label{rotbtzsolmu}
\eeq
to yield the analytic solution of $\mu$:
\beq
\mu=
\frac{2}{[(r_{+}-r_{1})(r_{-}-r_{2})]^{1/2}}[(r_{-}-r_{1})\Pi\left(c_{41},c_{42},c_{43}\right)+r_{1}F(c_{41},c_{43})]=G(r).
\label{btzsolmu22}
\eeq
Here $(c_{41},c_{42},c_{43})$ are given by
\bea
c_{41}&=&\sin^{-1}\left(\frac{(r-r_{-})(r_{+}-r_{1})}{(r-r_{1})(r_{+}-r_{-})}\right)^{1/2},\nn\\
c_{42}&=&\frac{r_{+}-r_{-}}{r_{+}-r_{1}},\nn\\
c_{43}&=&\left(\frac{(r_{+}-r_{-})(r_{1}-r_{2})}{(r_{+}-r_{1})(r_{-}-r_{2})}\right)^{1/2},
\eea
$\Pi(a,b,c)$ is the elliptic integral of the third kind defined in (\ref{piabc}) and $F(a,b)$ is given by (\ref{piabc22}).
One notes that $G(r)$ satisfies the boundary condition 
\beq
{\rm lim}_{r\rightarrow r_{-}}G(r)=0,
\label{btzbdy21}
\eeq
and $dr/d\mu >0$ implies that $G^{-1}$ is a well-defined function.  In the vanishing angular
momentum limit $J=0$, the above solution (\ref{btzsolmu22}) reduces to the static type-I extended BTZ 
spacetime case (\ref{btzsolmu}).

Exploiting the above coordinate (\ref{btzsolmu22}), we obtain
\beq
ds^{2}=-d\mu^{2}+f_{1}(\mu)^{2}dt^{2}+f_{2}^{2}(\mu)(d\phi+N^{\phi}dt)^{2}
\label{rotmetric}
\eeq
to yield the metric of the warped product form (\ref{btzmetric2}) in comoving coordinates where
one can replace~\cite{deser99,hong03} $d\phi+N^{\phi}dt\rightarrow d\phi$
to obtain the modified warping functions $f_{1}$ and $f_{2}$ as below
\bea
f_{1}(\mu)&=&\left(M-\frac{MB}{G^{-1}(\mu)}-\frac{(G^{-1}(\mu))^{2}}{l^{2}}-\frac{J^{2}}{4(G^{-1}(\mu))^{2}}\right)^{1/2},
\nonumber\\
f_{2}(\mu)&=&G^{-1}(\mu).
\label{rotbtzf1f2}
\eea
Here one notes that the detector
locates in the comoving coordinates with the angular velocity $d\phi/dt=-g_{t\phi}/g_{\phi\phi}=-N^{\phi}$
~\cite{deser99,hong03}.
We then arrive at the Ricci curvature components
\bea
R_{\mu\mu}&=&-\frac{2f_{1}^{\pr}}{f_{2}}+\frac{3MB}{2f_{2}^{3}}+\frac{J^{2}}{f_{2}^{4}},\nonumber\\
R_{tt}&=&\frac{2f_{1}^{2}f_{1}^{\pr}}{f_{2}}-\frac{3MBf_{1}^{2}}{2f_{2}^{3}}-\frac{J^{2}f_{1}^{2}}{f_{2}^{4}},\nonumber\\
R_{\phi\phi}&=&2f_{2}f_{1}^{\pr}.
\label{rotbtzriccis}
\eea
Here one notes that there does not exist an additional term associated with the angular
momentum $J$ in the $R_{\phi\phi}$ component since we have used the comoving coordinates.
The Einstein scalar curvature is then given by
\beq
R=-\frac{6}{l^{2}}-\frac{J^{2}}{2f_{2}^{4}},
\label{rotein}
\eeq
in the interior of the outer event horizon of the rotating type-I extended BTZ black hole.  Note that in the $J=0$ limit,
the above Ricci components (\ref{rotbtzriccis}) and Einstein scalar curvature (\ref{rotein})
reduce to the corresponding ones in (\ref{btzriccis}) and (\ref{btzeinr}) in the static type-I extended BTZ black hole, as 
expected.

%%%%%%%%%%%%%%%%%%%%%%%%%%%%%%%%%%%%%%%%%%%%%%%%%%%%%%%%%%%%%%%%%%%%%%
\section{Type-II extended BTZ spacetime}
\setcounter{equation}{0}
\renewcommand{\theequation}{\arabic{section}.\arabic{equation}}
%%%%%%%%%%%%%%%%%%%%%%%%%%%%%%%%%%%%%%%%%%%%%%%%%%%%%%%%%%%%%%%%%%%%%%
%%%%%%%%%%%%%%%%%%%%%%%%%%%%%%%%%%%%%%%%%%%%%%%%%%%%%%%%%%%%%%%%%%%%%%%
\subsection{Static case}
%%%%%%%%%%%%%%%%%%%%%%%%%%%%%%%%%%%%%%%%%%%%%%%%%%%%%%%%%%%%%%%%%%%%%%

Now, we consider an asymptotically constant 
scalar field case with the Lagrangian~\cite{chan97}
\beq
{\cal L}=\sqrt{-g}\left[\chi R-\frac{4\chi-1}{2\chi(1-\chi)}(\na \chi)^{2}+\frac{M}{2L}
+6\left(2\Lambda-\frac{M}{2L}\right)\chi
+18\left(-\Lambda+\frac{M}{4L}\right)\chi^{2}
+2\left(4\Lambda-\frac{M}{L}\right)\chi^{3}\right],
\label{lagl}
\eeq
where
\beq
\chi=\frac{r^{2}}{r^{2}-2L},
\eeq
which is associated with a metric of the form
\bea
ds^{2}&=&-\bar{N}^{2}dt^{2}+\bar{N}^{-2}dr^{2}+r^{2}d\phi^{2},\nn\\
\bar{N}^{2}&=&-M+\frac{ML}{r^{2}}+\frac{r^{2}}{l^{2}}.\label{dsbar2}
\eea 
Here $L$ in (\ref{dsbar2}) is again introduced to couple 
the scalar field $\chi$ to the metric $g_{\mu\nu}$ gravity. 
Moreover, $L$ in $\bar{N}^{2}$ of (\ref{dsbar2}) is trivially related to $L$ in 
the modified lapse function $N^{2}$ for the warp product scheme since $N^{2}=-\bar{N}^{2}$ in (\ref{btz2lapse}).
On the other hand, investigating the lapse functions $\bar{N}^{2}$ and $N^{2}$ (for warp product case), one 
readily observe that the only curvature singularity is located at $r=0$~\cite{chan97}. 
Moreover, in the limit $L=(J^{2}/4M)$, the metric seems to be related to a rotating 
BTZ black hole. However, because the three-metric (\ref{dsbar2}) does not contain a shift function, the 
metric does not allow such  a rotating BTZ black hole. The rotating 
type-II extended BTZ spacetime case will be investigated 
in Section III-B below. Variating the Lagrangian in (\ref{lagl}) with respect to the metric, 
we obtain the Einstein equation in (\ref{einstein}) with the energy-stress tensor given by
\bea
T_{\mu\nu}&=&\frac{4\chi-1}{2\chi^{2}(1-\chi)}\na_{\mu}\chi\na_{\nu}\chi+\frac{1}{2}g_{\mu\nu}
\left[-\frac{4\chi-1}{2\chi^{2}(1-\chi)}(\na \chi)^{2}
+\frac{M}{2L\chi}
\right.\nn\\
&&\left.+6\left(2\Lambda-\frac{M}{2L}\right)+18\left(-\Lambda+\frac{M}{4L}\right)\chi+2\left(4\Lambda-\frac{M}{L}\right)\chi^{2}\right].
\eea
Next, variation of the Lagrangian in (\ref{lagl}) with respect to the scalar field $\chi$ produces
\beq
R+\frac{1-2\chi+4\chi^{2}}{2\chi^{2}(1-\chi)^{2}}(\na\chi)^{2}+\frac{4\chi-1}{\chi(1-\chi)}\na^{2}\chi
+6\left(2\Lambda-\frac{M}{2L}\right)+36\left(-\Lambda+\frac{M}{4L}\right)\chi+6\left(4\Lambda-\frac{M}{L}\right)\chi^{2}=0.
\eeq 

Now we consider a multiply warped product manifold associated with the
static type-II extended BTZ spacetime inside the event horizon where three-metric (\ref{btzmetric})
is given in terms of the lapse function
\beq
N^{2}=M-\frac{ML}{r^{2}}-\frac{r^{2}}{l^{2}}.
\label{btz2lapse}
\eeq
The roots of $N^{2}=0$ for arbitrary real value of $r$ are given by
\bea
r_{+}&=&\left(\frac{Ml^{2}+(M^{2}l^{4}-4MLl^{2})^{1/2}}{2}\right)^{1/2},\nn\\
r_{-}&=&\left(\frac{Ml^{2}-(M^{2}l^{4}-4MLl^{2})^{1/2}}{2}\right)^{1/2}=ir_{1}.
\label{btz2rpmii}
\eea
We find an identity
\beq
r_{+}^{2}+r_{-}^{2}=Ml^{2}.\label{id20}
\eeq
For a given nonzero $L$, the mass parameter $M$ is given by the roots $r_{+}$ and $r_{-}$
\beq
M=\frac{r_{+}^{2}r_{-}^{2}}{l^{2}L}.
\eeq
For a vanishing $L$, we also obtain $M$ in (\ref{mbzero}) for the BTZ case.
The lapse function can be classified in terms of the value of $L$ and $r_{\pm}$.

(i) For $0<L<L_{1}$ with 
\beq
L_{1}=\frac{Ml^{2}}{4},\label{l1def}
\eeq
we find two positive real roots $r_{+}$ and $r_{-}$ with $r_{+}>r_{-}$ to produce
\beq
N^{2}=\frac{(r_{+}-r)(r-r_{-})(r+r_{+})(r+r_{-})}{l^{2}r^{2}},
\label{btz2lapserh1}
\eeq
which, for the interior solution, is well-defined in the region $r_{-}<r<r_{+}$.
Introducing the coordinate $\mu$ as in (\ref{btzdmu}) we obtain
\beq
\mu=\int_{r_{-}}^{r}dx~\left(\frac{l^{2}x^{2}}{(r_{+}-x)(x-r_{-})(x+r_{+})(x+r_{-})}\right)^{1/2},
\label{rotdssolmu}
\eeq
whose analytic solution is of the form
\beq
\mu=l\sin^{-1}\left(\frac{r^{2}-r_{-}^{2}}{r_{+}^{2}-r_{-}^{2}}\right)^{1/2}=G(r).
\label{btz2an}
\eeq
Here we have used the $r_{\pm}$ in (\ref{btz2rpmii}). We note that $G(r)$ satisfies the boundary condition (\ref{btzbdy21}) and 
$dr/d\mu >0$ implies that $G^{-1}$ is a well-defined function. 

(ii) For $L=0$, we have the BTZ case again. The ensuing argument is the same as that of (iii) of the static type-I, to yield the 
analytic solution (\ref{btzdmu13}) for $\mu$.

(iii) For $L<0$, $r_{+}$ and $r_{-}$ are a positive real root and an imaginary one, respectively. The lapse function is then 
given by
\beq
N^{2}=\frac{(r_{+}-r)(r+r_{+})(r^{2}+r_{1}^{2})}{l^{2}r^{2}},
\label{btz2lapserh12}
\eeq
which, for the interior solution, is well-defined in the region $0<r<r_{+}$. We introduce 
the coordinate $\mu$ as in (\ref{btzdmu}) to obtain
\beq
\mu=\int_{0}^{r}dx~\left(\frac{l^{2}x^{2}}{(r_{+}-x)(x+r_{+})(x^{2}+r_{1}^{2})}\right)^{1/2}=G(r).
\label{rotdssolmu3}
\eeq
We note that $G(r)$ satisfies the boundary condition (\ref{btzbdy3}) and 
$dr/d\mu >0$ implies that $G^{-1}$ is a well-defined function. If a root of $N^{2}=0$ does not belong to the above 
categories, we cannot obtain the coordinate $\mu$.

Exploiting the above coordinate (\ref{btzdmu13}), (\ref{btz2an}) and (\ref{rotdssolmu3}), we can obtain the warped products (\ref{btzmetric2}) and the 
corresponding modified warping functions $f_{1}$ and $f_{2}$ given as below
\bea
f_{1}(\mu)&=&\left(M-\frac{ML}{(G^{-1}(\mu))^{2}}-\frac{(G^{-1}(\mu))^{2}}{l^{2}}\right)^{1/2},
\nonumber\\
f_{2}(\mu)&=&G^{-1}(\mu),
\label{btz2f1f2}
\eea
to arrive at, in the interior of the outer event horizon of the static type-II extended BTZ spacetime, the
Ricci curvature components, 
\bea
R_{\mu\mu}&=&-\frac{2f_{1}^{\pr}}{f_{2}}+\frac{4ML}{f_{2}^{4}},\nonumber\\
R_{tt}&=&\frac{2f_{1}^{2}f_{1}^{\pr}}{f_{2}}-\frac{4MLf_{1}^{2}}{f_{2}^{4}},\nonumber\\
R_{\phi\phi}&=&2f_{2}f_{1}^{\pr}
\label{btz2riccis}
\eea
and the Einstein scalar curvature
\beq
R=-\frac{6}{l^{2}}-\frac{2ML}{f_{2}^{4}}.
\label{btz2inr}
\eeq

%%%%%%%%%%%%%%%%%%%%%%%%%%%%%%%%%%%%%%%%%%%%%%%%%%%%%%%%%%%%%%%%%%%%%%%
\subsection{Rotating case}
%%%%%%%%%%%%%%%%%%%%%%%%%%%%%%%%%%%%%%%%%%%%%%%%%%%%%%%%%%%%%%%%%%%%%%

Next we consider a multiply warped product manifold associated with the
rotating type-II extended BTZ spacetime inside the event horizon where three-metric (\ref{rotbtzmetric})
is given by the lapse and shift functions:
\bea
N^{2}&=&M-\frac{ML}{r^{2}}-\frac{r^{2}}{l^{2}}-\frac{J^{2}}{4r^{2}},
\nonumber\\
N^{\phi}&=&-\frac{J}{2r^{2}}.
\label{rotdslapse}
\eea
The roots of $N^{2}=0$ for arbitrary real value of $r$ are given by
\bea
r_{+}&=&\left(\frac{Ml^{2}+(M^{2}l^{4}-4MLl^{2}-J^{2}l^{2})^{1/2}}{2}\right)^{1/2},\nn\\
r_{-}&=&\left(\frac{Ml^{2}-(M^{2}l^{4}-4MLl^{2}-J^{2}l^{2})^{1/2}}{2}\right)^{1/2}=ir_{2}.
\label{btz2rpmii2}
\eea
We also obtain the identity (\ref{id20}). For a given nonzero $L$, the mass parameter $M$ is given in terms 
of the roots and the angular momentum
\beq
M=\frac{r_{+}^{2}r_{-}^{2}-\frac{1}{4}J^{2}l^{2}}{l^{2}L}.
\eeq
We also obtain the identity (\ref{id20}).
The lapse function can be classified in terms of the value of $L+\frac{J^{2}}{4M}$ and $r_{\pm}$.

(i) For $0<L+\frac{J^{2}}{4M}<L_{1}$ with $L_{1}$ in (\ref{l1def}), 
we find two positive real roots $r_{+}$ and $r_{-}$ with $r_{+}>r_{-}$ to yield the lapse function of 
the form (\ref{btz2lapserh12}) which, for the interior solution, is well-defined in the region $r_{-}<r<r_{+}$.
Introducing the coordinate $\mu$ as in (\ref{btzdmu}) we obtain of the coordinate $\mu$ in (\ref{rotdssolmu}).
Here we note that $r_{+}$ and $r_{-}$ are different from that of the static case since we have a rotation term in 
(\ref{btz2rpmii2}). The analytic solution of (\ref{rotdssolmu}) is given by
\beq
\mu=l\sin^{-1}\left(\frac{r^{2}-r_{-}^{2}}{r_{+}^{2}-r_{-}^{2}}\right)^{1/2}=G(r).
\label{btz2an21}
\eeq
Here we have used the $r_{\pm}$ in (\ref{btz2rpmii2}). We note that $G(r)$ satisfies the boundary condition (\ref{btzbdy21}) and 
$dr/d\mu >0$ implies that $G^{-1}$ is a well-defined function. 

(ii) For $L+\frac{J^{2}}{4M}=0$, we obtain the BTZ case. The ensuing argument is again the same as that of (iii) of the static type-I, 
to produce the analytic solution (\ref{btzdmu13}) for $\mu$ in the comoving coordinates.

(iii) For $L+\frac{J^{2}}{4M}<0$, $r_{+}$ and $r_{-}$ are a positive real root and an imaginary one, respectively. The lapse function is now given by
\beq
N^{2}=\frac{(r_{+}-r)(r+r_{+})(r^{2}+r_{2}^{2})}{l^{2}r^{2}},
\label{btz2lapserh13}
\eeq
which, for the interior solution, is well-defined in the region $0<r<r_{+}$. We introduce 
the coordinate $\mu$ as in (\ref{btzdmu}) to obtain
\beq
\mu=\int_{0}^{r}dx~\left(\frac{l^{2}x^{2}}{(r_{+}-x)(x+r_{+})(x^{2}+r_{2}^{2})}\right)^{1/2}=G(r).
\label{rotdssolmu33}
\eeq
We note that $G(r)$ satisfies the boundary condition (\ref{btzbdy3}) and 
$dr/d\mu >0$ implies that $G^{-1}$ is a well-defined function. If a root of $N^{2}=0$ does not belong to the above 
categories, we cannot construct the coordinate $\mu$.

Exploiting the above coordinate (\ref{btzdmu13}), (\ref{btz2an21}),  and (\ref{rotdssolmu33}), we obtain the metric (\ref{rotmetric})
and then we find the warped products (\ref{btzmetric2}) in the comoving coordinates where
one can replace $d\phi+N^{\phi}dt\rightarrow d\phi$ and the modified warping functions $f_{1}$ and
$f_{2}$ are given as below
\bea
f_{1}(\mu)&=&\left(M-\frac{ML}{(G^{-1}(\mu))^{2}}-\frac{(G^{-1}(\mu))^{2}}{l^{2}}-\frac{J^{2}}{4(G^{-1}(\mu))^{2}}\right)^{1/2},
\nonumber\\
f_{2}(\mu)&=&G^{-1}(\mu),
\label{rotbtz2f1f2}
\eea
to yield, in the interior of the outer event horizon of the rotating type-II extended BTZ spacetime, the
Ricci curvature components, 
\bea
R_{\mu\mu}&=&-\frac{2f_{1}^{\pr}}{f_{2}}+\frac{4ML}{f_{2}^{4}}+\frac{J^{2}}{f_{2}^{4}},\nonumber\\
R_{tt}&=&\frac{2f_{1}^{2}f_{1}^{\pr}}{f_{2}}-\frac{4MLf_{1}^{2}}{f_{2}^{4}}-\frac{J^{2}f_{1}^{2}}{f_{2}^{4}},\nonumber\\
R_{\phi\phi}&=&2f_{2}f_{1}^{\pr}
\label{rotbtz2riccis}
\eea
and the Einstein scalar curvature
\beq
R=-\frac{6}{l^{2}}-\frac{2ML}{f_{2}^{4}}-\frac{J^{2}}{2f_{2}^{4}}.
\label{rotbtz2inr}
\eeq
Note that in the $J=0$ limit, the above Einstein scalar curvature (\ref{rotbtz2inr})
reduces to the corresponding one (\ref{btz2inr}) in the static type-II extended BTZ spacetime case.

%%%%%%%%%%%%%%%%%%%%%%%%%%%%%%%%%%%%%%%%%%%%%%%%%%%%%%%%%%%%%%%%%%%%%%%
\section{Conclusions}
\setcounter{equation}{0}
\renewcommand{\theequation}{\arabic{section}.\arabic{equation}}
\label{sec:conclusions}
%%%%%%%%%%%%%%%%%%%%%%%%%%%%%%%%%%%%%%%%%%%%%%%%%%%%%%%%%%%%%%%%%%%%%%

We have studied the multiply warped product manifold associated with the ST theories of the BTZ
black holes to evaluate the Ricci curvature components inside the black hole event horizons. Exploiting these
Ricci curvatures, we have shown that all the Einstein scalar curvatures are 
identical both in the exterior and
interior of the outer event horizons without discontinuities for these extended BTZ spacetimes. 

To investigate the ST theories, we have considered two cases with $c(\chi)=\chi$ in (\ref{action0}) where gravitational forces are given by a mixture of the metric and the scalar field. Here the scalar field is an ordinary one, not a phantom field. 
Using the Lagrangian for the metric and scalar field, we have investigated the Einstein equations and warp product aspects of the ST theories. To do this we have exploited the algorithm for roots of cubic and quartic equations 
associated with the warp products.

\acknowledgments 
The author would like to thank the anonymous referees for helpful comments.

%%%%%%%%%%%%%%%%%%%%%%%%%%%%%%%%%%%%%%%%%%%%%%%%%%%%%%%%%%%%%%%%%%%%%%%%%
\appendix
\section{Mathematical aspects of a quartic equation}
\setcounter{equation}{0}
\renewcommand{\theequation}{A.\arabic{equation}}
%%%%%%%%%%%%%%%%%%%%%%%%%%%%%%%%%%%%%%%%%%%%%%%%%%%%%%%%%%%%%%%%%%%%%%%%%

Now, we consider four roots of the lapse function in (\ref{rotbtzlapse1}) which produces a quartic equation of the form
\beq
x^{4}+a_{1}x^{2}+a_{2}x+a_{3}=0,
\label{quarticeqn}
\eeq
where
\beq
a_{1}=-Ml^{2},~~~a_{2}=MBl^{2},~~~a_{3}=\frac{1}{4}J^{2}l^{2}.
\label{a1a2a3}
\eeq
Following the algorithm for treating the quartic equation~\cite{weisstein03}, we obtain the four roots 
$(x_{1},x_{2},x_{3},x_{4})$ of the equation (\ref{quarticeqn}) given by
\bea
x_{1}&=&\frac{1}{2}p_{1}+\frac{1}{2}p_{2},\nn\\
x_{2}&=&\frac{1}{2}p_{1}-\frac{1}{2}p_{2},\nn\\
x_{3}&=&-\frac{1}{2}p_{1}+\frac{1}{2}p_{3},\nn\\
x_{4}&=&-\frac{1}{2}p_{1}-\frac{1}{2}p_{3},\label{xxxx}
\eea 
where 
\beq
p_{1}=\left(y_{0}-a_{1}\right)^{1/2}.
\eeq
For $p_{1}\neq 0$ we obtain~\footnote{In the literature~\cite{weisstein03}, there exist typos in the expressions for $p_{2}$ 
in (\ref{peneq}), and for $q_{1}$ and $q_{2}$ in (\ref{qqqq}).}
\bea
p_{2}&=&\left(-p_{1}^{2}-2a_{1}-\frac{2a_{2}}{p_{1}}\right)^{1/2},\nn\\
p_{3}&=&\left(-p_{1}^{2}-2a_{1}+\frac{2a_{2}}{p_{1}}\right)^{1/2},\label{peneq}
\eea
and for $p_{1}=0$ we arrive at
\bea
p_{2}&=&\left(-2a_{1}+2\left(y_{0}^{2}-4a_{3}\right)^{1/2}\right)^{1/2},\nn\\
p_{3}&=&\left(-2a_{1}-2\left(y_{0}^{2}-4a_{3}\right)^{1/2}\right)^{1/2}.
\eea 
Here $y_{0}$ is defined as a real root of the following cubic equation
\beq
y^{3}+b_{1}y^{2}+b_{2}y+b_{3}=0,
\label{cubiceq}
\eeq
where
\beq
b_{1}=-a_{1},~~~b_{2}=-4a_{3},~~~b_{3}=4a_{1}a_{3}- a_{2}^{2}.
\eeq
The three roots of the cubic equation (\ref{cubiceq}) are then readily given by (see Appendix B for details.)
\bea
y_{1}&=&q_{1}+q_{2}-\frac{1}{3}b_{1},\nn\\
y_{2}&=&-\frac{1}{2}(q_{1}+q_{2})+\frac{i\sqrt{3}}{2}(q_{1}-q_{2})-\frac{1}{3}b_{1},\nn\\
y_{3}&=&-\frac{1}{2}(q_{1}+q_{2})-\frac{i\sqrt{3}}{2}(q_{1}-q_{2})-\frac{1}{3}b_{1},
\label{yyy}
\eea
where
\bea
q_{1}&=&\left(q_{3}+(q_{3}^{2}+q_{4}^{3})^{1/2}\right)^{1/3},\nn\\
q_{2}&=&\left(q_{3}-(q_{3}^{2}+q_{4}^{3})^{1/2}\right)^{1/3},\nn\\
q_{3}&=&\frac{9b_{1}b_{2}-27b_{3}-2b_{1}^{2}}{54},\nn\\
q_{4}&=&\frac{3b_{2}-b_{1}^{2}}{9}.\label{qqqq}
\eea
These three roots in (\ref{yyy}) can be also rewritten in terms of trigonometric functions as follows 
\bea
y_{1}&=&2(-q_{4})^{1/2}\cos\left(\frac{\psi}{3}\right)-\frac{1}{3}b_{1},\nn\\
y_{2}&=&2(-q_{4})^{1/2}\cos\left(\frac{\psi}{3}+\frac{2\pi}{3}\right)-\frac{1}{3}b_{1},\nn\\
y_{3}&=&2(-q_{4})^{1/2}\cos\left(\frac{\psi}{3}+\frac{4\pi}{3}\right)-\frac{1}{3}b_{1},
\label{yyytri}
\eea
where $\psi$ satisfies the identities
\bea
\cos\psi&=&\frac{q_{3}}{(-q_{4}^{3})^{1/2}},\nn\\
\sin\psi&=&\left(1+\frac{q_{3}^{2}}{q_{4}^{3}}\right)^{1/2}.\label{psi}
\eea

Now, we consider the specific case of the static BTZ limit where the coefficients of the quartic equation (\ref{quarticeqn}) are given as
\beq
a_{1}=-Ml^{2},~~~a_{2}=0,~~~a_{3}=0.
\eeq
From (\ref{yyy}), one readily obtain the three roots of the cubic equation (\ref{cubiceq})
\beq
y_{1}=a_{1},~~~y_{2}=0,~~~y_{3}=0.
\label{100}
\eeq
If we choose $y_{0}$ as in $y_{0}=y_{1}=a_{1}=-Ml^{2}$, exploiting (\ref{xxxx}) we obtain the four roots of the quartic equation (\ref{quarticeqn}) as follows
\beq
x_{1}=0,~~~
x_{2}=0,~~~
x_{3}=M^{1/2}l,~~~
x_{4}=-M^{1/2}l,\label{xxxxstatic1}
\eeq
which imply that $r_{+}=x_{3}$, $r_{-}=x_{4}$, $r_{1}=x_{1}$ and $r_{2}=x_{2}$.
Similarly for the choice of $y_{0}=y_{2}=y_{3}=0$, we again obtain the four roots of the equation (\ref{quarticeqn}) 
\beq
x_{1}=M^{1/2}l,~~~
x_{2}=0,~~~
x_{3}=0,~~~
x_{4}=-M^{1/2}l,\label{xxxxstatic2}
\eeq
to conclude that $r_{+}=x_{1}$, $r_{-}=x_{4}$, $r_{1}=x_{2}$ and $r_{2}=x_{3}$. The mapping of $(r_{+},r_{-},r_{1},r_{2})$ 
onto $(x_{1},x_{2},x_{3},x_{4})$ depends on the choice of $y_{0}$ in $(y_{1},y_{2},y_{3})$.

%%%%%%%%%%%%%%%%%%%%%%%%%%%%%%%%%%%%%%%%%%%%%%%%%%%%%%%%%%%%%%%%%%%%%%%%%
%\appendix
\section{Algorithm for roots of a cubic equation}
\setcounter{equation}{0}
\renewcommand{\theequation}{B.\arabic{equation}}
%%%%%%%%%%%%%%%%%%%%%%%%%%%%%%%%%%%%%%%%%%%%%%%%%%%%%%%%%%%%%%%%%%%%%%%%%

Next, we recapitulate the algorithm for finding solutions of a cubic equation
appeared in~\cite{weisstein03}. To do this, we start with a cubic equation of the form
\beq
y^{3}+b_{1}y^{2}+b_{2}y+b_{3}=0.
\label{y3}
\eeq 
Introducing a new variable
\beq
z=y+\frac{1}{3}b_{1},
\label{zy}
\eeq
we find the following form
\beq
z^{3}+3q_{4}z-2q_{3}=0,
\label{z3eqn}
\eeq
where $q_{3}$ and $q_{4}$ are given by (\ref{qqqq}).
Next, we readily check that 
\beq
z_{1}=q_{1}+q_{2},
\label{q1q2}
\eeq
with $q_{1}$ and $q_{2}$ being defined as in (\ref{qqqq}), is a root of the cubic equation (\ref{z3eqn}).
Moreover, we find that (\ref{z3eqn}) is factorized as 
\beq
(z-z_{1})(z^{2}+z_{1}z+3q_{4})=0,
\eeq
from which we obtain the other two roots of the from
\bea
z_{2}&=&-\frac{1}{2}(q_{1}+q_{2})+\frac{i\sqrt{3}}{2}(q_{1}-q_{2}),\nn\\
z_{3}&=&-\frac{1}{2}(q_{1}+q_{2})-\frac{i\sqrt{3}}{2}(q_{1}-q_{2}).
\label{z2z3}
\eea
Even though $z_{2}$ and $z_{3}$ possess an $i$ as shown in (\ref{z2z3}), this does not 
indicate anything about the numbers of real and complex roots, since $q_{1}$ and $q_{2}$ are themselves 
complex in general. In order to determine which roots are real or complex, we need to introduce 
a criterion parameter, namely the discriminant $D$ defined as
\beq      
D=q_{3}^{2}+q_{4}^{3}.
\eeq
If $D>0$, we have one real root and two complex conjugates; if $D=0$, we have three real roots and 
at least two roots are equal; if $D<0$, we have three real roots and all roots are different. 
For our case of interest associated with (\ref{rotlapserh}), we need to have three different roots in the cubic 
equation. From now on, we will thus consider $D<0$ case only. Keeping these roots $(z_{1},z_{2},z_{3})$ 
in mind and using the definition (\ref{zy}), 
we readily find the roots (\ref{yyy}) for the cubic equation in (\ref{cubiceq}) or in (\ref{y3}).
Moreover, we obtain the identities:
\bea
z_{1}+z_{2}+z_{3}&=&0,\nn\\
z_{1}z_{2}+z_{1}z_{3}+z_{2}z_{3}&=&3q_{4},\nn\\
z_{1}z_{2}z_{3}&=&2q_{3},\nn\\
z_{1}^{2}+z_{2}^{2}+z_{3}^{2}&=&-6q_{4},\nn\\
z_{1}^{3}+z_{2}^{3}+z_{3}^{3}&=&6q_{3},\nn\\
z_{1}^{4}+z_{2}^{4}+z_{3}^{4}&=&18q_{4}^{2},\nn\\
z_{1}^{5}+z_{2}^{5}+z_{3}^{5}&=&-30q_{3}q_{4}.
\eea

Finally, we formulate the above roots in terms of trigonometric function. To do this, we exploit the angle 
$\psi$ defined in (\ref{psi}). Here we again consider the negative discriminant case: $D<0$ which is relevant to our 
cubic equation at hand. Exploiting the identities
\beq
q_{3}\pm i(-q_{3}^{2}-q_{4}^{3})^{1/2}=(-q_{4}^{3})^{1/2}e^{\pm i\psi},
\eeq 
we arrive at the desired forms in (\ref{yyytri}).

\end{document}